\documentclass{article}
\usepackage{graphicx}
\usepackage{changepage}
\begin{document}
\title{Market Formation as Transitive Closure: the Evolving Pattern of Trade in Music\thanks{accepted for publication in \textit{Network Science}}}
\author{Jesse Shore\\
		Boston University\\
		jccs@bu.edu}
\date{\today}
%%%%%%%%%%%%%%%%

%%%%%%%%%%%%%%%%

\maketitle % Insert title

%----------------------------------------------------------------------------------------
%	ABSTRACT
%----------------------------------------------------------------------------------------

\begin{abstract}
\noindent  % Dummy abstract text
Where do new markets come from?  I construct a network model in which national markets are nodes and flows of recorded music between them are links and conduct a longitudinal analysis of the global pattern of trade in the period 1976 to 2010.  I hypothesize that new export markets are developed through a process of transitive closure in the network of international trade.  When two countries' markets experience the same social influences, it brings them close enough together for new homophilous ties to be formed.  The implication is that consumption of foreign products helps, not hurts, home-market producers develop overseas markets, but only in those countries that have a history of consuming the same foreign products that were consumed in the home market. Selling in a market changes what is valued in that market, and new market formation is a consequence of having social influences in common.
\end{abstract}

\section{Introduction}

Many innovative industries produce goods or services that have an interpreted, or ``cultural'' component to the value placed on them by consumers, in that tastes differ in different markets: just because a product is ``big in Japan'' does not mean it is ``big'' elsewhere. The same could be said at the level of national industries: as Figure 1  illustrates, for example, although there may be a market for my home country's music in Japan, there may not be one in Italy. We are told that the division of labor is limited by the extent of the market, but for information goods the more fundamental issue may be what the extent of the market is limited by: why do only certain other countries have substantial demand for my industry's innovative products?  

The question has substantial practical consequences.  For individual producers and distributors of information goods, selling to foreign markets can be highly profitable due to negligible marginal costs for information goods (Wildman and Siwek 1988, Parker and Van Alstyne, 2005); for whole economies, the presence of high-value export industries can drive overall growth and development (Amsden, 2001; Hausmann, Hwang, and Rodrik, 2007; Hidalgo et al, 2007).  Despite the importance of the question, however, received economic theory does not provide a clear answer.  

\begin{figure}[ht]	
\label{fig:IVJ}
	\centering
		\includegraphics[width=1.00\textwidth]{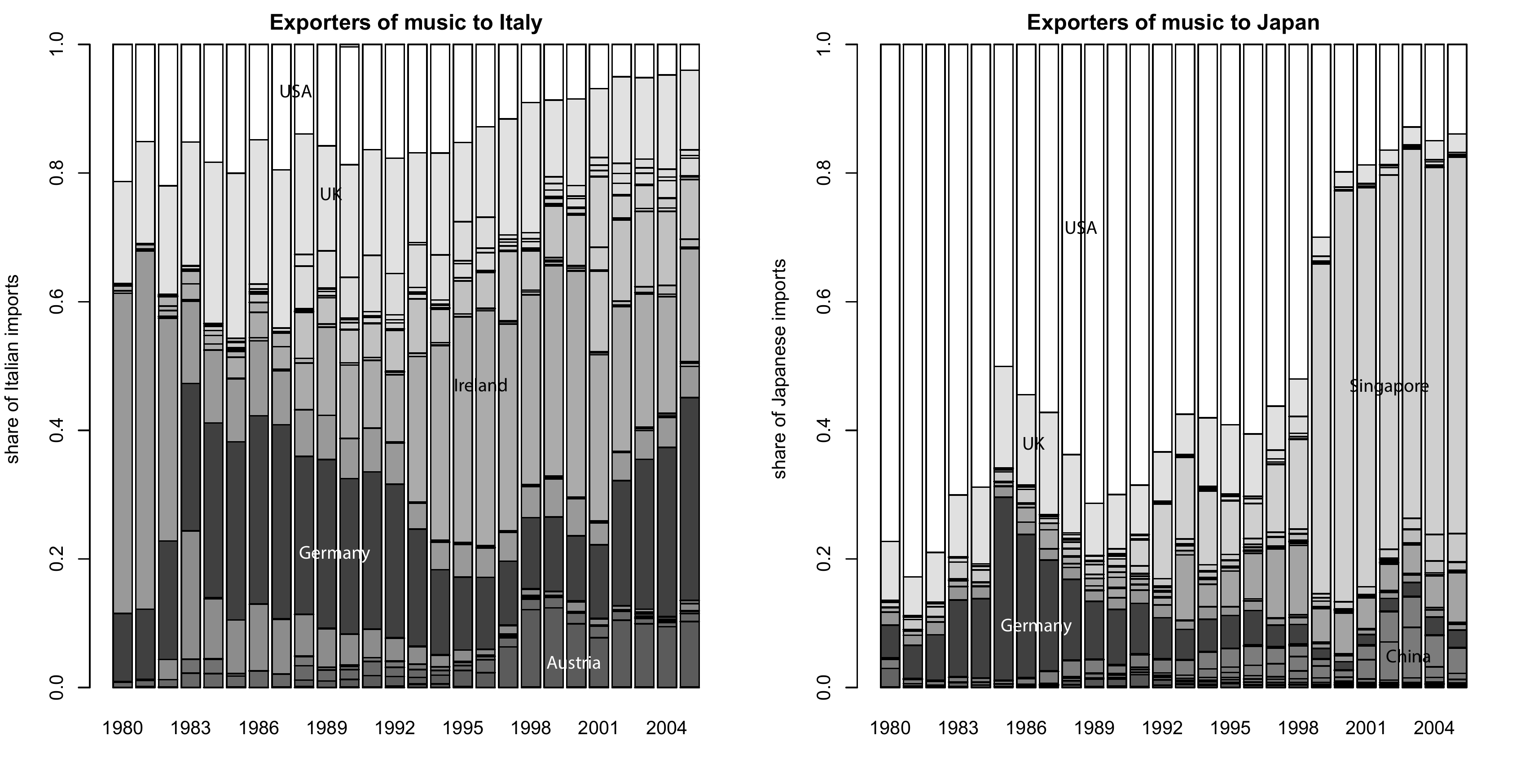}
	\caption{Demand for interpreted goods is not uniform across countries.  For example, the composition of Italian imports of recorded music (left) is strikingly different from the composition of Japan's (right)}

\end{figure}

The economic viewpoint is that specialization promotes exchange, and that specialization must be maintained to remain competitive (Ricardo, 1891, Krugman, 1980, 1991; Storper and Christopherson, 1987; Storper, 1992; Malmberg and Maskell, 1997).  By this logic, anything that would reduce the distinctiveness of a country's products would diminish its ability to export them.  Accordingly, there is some fear in certain quarters that homogenizing effects of consuming imports -- especially of American goods --  would lessen the distinctiveness and global profitability of many countries' information industries (Henry, 2001; World Trade Organization Council for Trade in Services, 1998; UNESCO, 2005; African Union, 2005).  

Is this a well-founded fear?  If distinctiveness is an asset for national information industries, is it better for them to be relatively insulated from imports? Or is it actually better to be interconnected and influenced by the outside? The opportunities and threats are greatest for smaller industries, in which potential outcomes for individual producers and distributors range from entering more lucrative foreign markets to being crowded out by global superstars. But understanding the dynamic effects of cross-market trade would also pose substantial strategic questions for multi-national enterprises as well.  If Warner Music wants to expand its sales of its Belgian intellectual property in the United States, for example, how much of their American or British repertoire should they license to distribute on online music streaming services in Belgium?  Would too much outside influence dilute or diminish the distinctive 'Belgianness' and lessen the export value of Belgian repertoire? Or, on the contrary, would such influence make the music more familiar and valuable to new markets abroad?  More generally, where do new markets for interpreted information goods come from?

Although mainstream economic thought does not present a clear answer, the very term 'cultural good' suggests an alternative framing.  Cultural goods are 'cultural' in that they derive their value from subjective, but socially contingent, meanings.  The export of cultural goods from one country to another can be seen as a form of international communication, in which the goods are the messages.  The meanings and value of cultural goods (as with any messages) are not universal or intrinsic, but depend instead on the language or expectations of the message's recipients for interpretation.

Therefore, I view this not as a problem of the functioning of a market, but rather as a macroscopic network constituted by semiotically valent exchanges.  In the model I propose, national markets are nodes and flows of information goods between them are links. I conduct a longitudinal network analysis of the global pattern of trade in a specific information industry --- recorded music --- in the period 1976 to 2010.  

Using this framing, the following sections develop and test the hypothesis that new export markets are developed through a process of transitive closure in the network of international trade.  When two countries' markets experience the same social influences, it brings them close enough together for new homophilous ties to be formed. In other words, consumption of foreign products at home helps, not hurts, home-market producers to enter new foreign markets, with the caveat that it only helps them enter markets that have a history of consuming the same foreign products that were consumed in the home market. Selling in a market changes what is valued in that market, and new market development can be seen as a consequence of having social influences in common. 

\section{Social influence in inter-market trade}
In general, it is believed that people are more ``homogenous within than between groups'' (Burt 2004), and that explanations for this pattern can be grouped into two broad types: homophily and social influence (c.f. Aral, Muchnik, and Sundararajan, 2009).  These two explanations are not mutually exclusive, but explain complementary phenomena.  Homophily is the tendency for individuals to form ties with others like themselves.  Social influence, on the other hand, suggests that connected individuals will become more alike. Both have salient interpretations in the present context.  A tie suggests that the products of the exporting country are intelligible enough to consumers in the importing country to support a relatively high volume of trade.  Indeed, by the very act of modeling international trade as a network of meaningful exchange ties, homophily is assumed by construction.  In seeking to understand how the pattern of trade \textit{changes over time}, however, we are interested in the role that social influence plays in concert with homophily.  

Importantly, in the context of music trade, the primary vehicle for influence is the music recording, rather than interpersonal contact with individuals in other countries.  Music is a form of symbolic communication, so the export of music from one country to another amounts to a sort of macroscopic communication tie.  Although music listeners in the importing country typically do not directly interact with music listeners in the exporting country, they are on the receiving end of communicative exchange, and may be influenced by listening to a subset of the exporting country's music. 

Messages transmitted from one person to another in the process of communication, whether linguistic or non-linguistic, have no intrinsic meaning but must be interpreted as corresponding to meanings by their recipients (e.g. Maines 1977, Lakoff 1987, Carlile, 2002). The sounds that make up a word in one language may be nonsense in another language, for example.  There is a requirement, therefore, for a learned interpretive layer --- a sort of semantic knowledge that is an aspect of language --- that maps messages to meanings: a 'tacit knowledge' (Polanyi 1966, Polanyi and Prosch 1977) or absorptive capacity (Cohen and Levinthal, 1990) on the demand side.   

To illustrate, consider a hypothetical repertoire of music. If musical phrases in this body of work conventionally end in a certain way, a songwriter could play with the listener's learned expectation that a specific phrase will end the same way. For example, by withholding the expected harmony and prolonging the penultimate chord, one might increase the dramatic tension. A piece of music that includes this technique depends for effect in part on the listener expecting the final harmony, only to have that expectation manipulated by the withholding of that harmony. If such music were brought to listeners from a different country who did not ``speak the same musical language,'' a long penultimate chord would not have any particular meaning because there would be no expectation in mind of what was supposed to happen. Therefore a musical innovation - in this case the displacement of the expected harmony in time - does not create an increase in quality by creating objectively better combinations of musical elements, but rather depends on the listener's expectations to achieve an effect. Over time, new and innovative music, in turn, becomes a reference point for composition and consumption of future music. To continue with the example of the withheld harmony, if it were used and heard enough, it could eventually become a conventional element in a phrase ending, and its effect on listeners would not in that case include surprise or increased tension.

This is not to say that every individual would have the same reaction; people respond to music differently. In the case of the withheld harmony, though everyone may have learned to expect the same conventional phrase endings, some subset of the population may prefer those expectations to be met.  These individuals would favor a more conventional style, while others may enjoy the innovation. Although their tastes differ, the fact that they share that reference point means that each other's preferred music is nonetheless mutually intelligible.  To people in other markets who are not familiar with the conventional phrase endings, there would be no difference in effect if the final harmony was withheld or not.  

Looking at music through this lens, what exposure to the same music would provide, therefore, is not a predictable influence on tastes that pertains to all listeners, but rather a common reference point, to which people may relate differently. To the degree that words are used in shared experiential contexts, they should contribute to establishing shared meanings among the people that use them to communicate.  However, to the extent that contexts and associations differ, the meanings associated with words will tend to diverge.  This suggests that shared meaning -- and perhaps shared value -- is possible only to the extent that the communicators have shared experience.  In summary, I have argued thus far that being exposed to music in common is a precondition for sharing aspects of a ``musical language,'' which is to say the set of expectations through which new music is understood and evaluated.

\subsection{Economic literature}
The economic literature has not focused on trade in cultural goods, but does nonetheless include several important, if peripheral, treatments of the topic. These contributions do not constitute a coherent perspective, but can be integrated by interpretation through the above framework and deserve consideration here.

Linder (1961) suggested that trade propensities among countries should be understood as forming a matrix, representing the degree to which they produced products appropriate to each other's aggregate tastes. In more recent studies of trade in cultural goods, a minority but significant perspective on the difficulty in developing new export markets revolves around empirical evidence for the idea of a ``cultural discount,'' goods from abroad are not valued as highly as domestic goods (Hoskins, Mirus, and Rozeboom 1989;  Huang, 2007; Lee, 2006; Park, 2006; Rauch and Trindade, 2009). These findings are entirely consistent with Linder's hypothesis: tastes differ, and they are generally thought to be related to a variety of factors more or less exogenous to the phenomenon of trade itself.

Others, especially Schulze (1999), go on to consider how tastes change, and how trade might itself affect tastes. One factor in changing tastes is a variety of network externality, in that consumers are thought to derive more value out of listening to music, the more people they have to share the experience with (c.f. Salganik, Dodds, and Watts 2006, Salganik and Watts 2009, who describe a short-run version of a similar phenomenon in terms of social influence).  Schultz (1999) uses the term ``consumption capital'' to describe tastes formed through experience, explaining that ``eventually, the reinforcing effect of consumption capital accumulation makes it a part of the national culture. This implies that trade in art should be a positive function of cultural proximity and that current trade is a positive function of past trade.''

\begin{figure}
	\centering
		\includegraphics[width=0.40\textwidth]{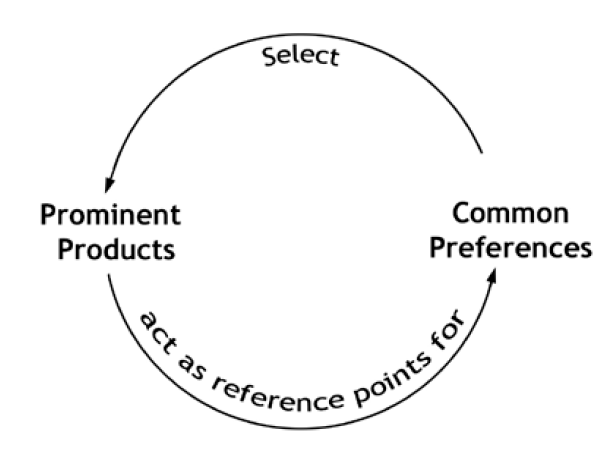}
	\caption{A co-evolutionary cycle between the characteristics of products and the preferences of consumers within a single market.}
	\label{fig:cycle}
\end{figure}

Witt (2001) expresses a similar argument in evolutionary terms: what has been consumed in the past conditions what is valued and consumed in the future. The conditioning is by means of the associations that the mind draws consciously or subconsciously between the old and the new. For higher-level and specific manifestations of wants (that is, those that are not innate such as food and shelter), it is by these ``learned associations'' that tastes become specific to certain products and types of products.  As tastes are conditioned by consumption they in turn make up the selection mechanism by which products gain in prominence and ability to further influence tastes (Figure \ref{fig:cycle}).

These minority perspectives in economic theory are clearly congruent with the reading of the sociologies of communication I advance above.  The task of the following section is to convert this perspective into a testable hypothesis.

\subsection{Hypothesis}
\begin{figure}
	\centering
		\includegraphics[width=0.50\textwidth]{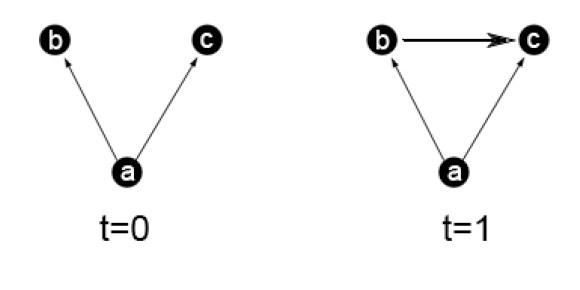}
	\caption{Hypothesized transitive closure due to social influence in common}
	\label{fig:TMTs}
\end{figure}

If importing music influences the local musical language, how should we expect this to alter the observed pattern of trade?  The most obvious implication would be toward a pattern of reciprocal trade.  If country A influences country B's musical language by exporting to it, the most obvious new market for country B's new output would be country A, because the two countries would have reference points in common.  Empirically, such a tendency to reciprocating trade ties would not be possible to attribute to the social influence-based mechanism I propose: reciprocal trade could be explained by homophily alone (which, as noted above, we assume must be present) or other dyadic latent variable without reference to social influence.  Latent variables are always problematic in empirical work, but they are a particular concern here.  If we start with a pair of countries that trade with each other, we know that they are already similar enough to support one homophilous tie, and there would be no need to invoke social influence in explaining why a second, reciprocated tie would be formed.  I suggest that looking beyond bilateral relationships would lessen these empirical difficulties, and that rather than look for exports back to the original exporter, we should look for exports to other countries that were also influenced by the original exporter (Figure \ref{fig:TMTs}).

To illustrate, consider a simple invented example. Imagine some country like the United States exports music to two other countries, say, Thailand and Nigeria. Some Thai producers and some Nigerian producers will be influenced by imports from the US, and some will not. However, assuming that a large proportion of consumers in those countries have consumed American music, we can expect the local musical language to be influenced by American music. If this process is happening in both Nigeria and Thailand simultaneously, we would expect their domestic musical output to be more intelligible to each other in the aggregate than before the influence of the United States. Given such an influence in common, we would therefore expect a subset of producers in Nigeria to have an easier time selling their products to Thai consumers (and vice versa) because of these shared reference points. If consumer tastes are also influenced by imports, in other words, then there should be an increased likelihood not just of exporting after importing, but of exporting to \emph{certain countries }that have common international influences from trade.

In the three country case, there are initially trade flows A $\rightarrow$ B and A $\rightarrow$ C, but no trade flows between B and C. There is therefore no prior evidence from the presence of trade that B and C have a high propensity to trade with each other, so a new trade flow B $\rightarrow$ C or C $\rightarrow$ B could be considered a more meaningful finding. There is admittedly still a selection bias, because B and C both import from A, but since it is more indirect, it can be assumed to be lesser than the selection bias inherent in looking for reciprocal trade. The tri-lateral approach is also advantageous in regard to effective sample size. If, say, the United States exported to n=30 countries, there are 30 possible reciprocal trade flows one could look for, but n!/(2!(n- 2)!) = 435 cases of triadic closure we could look for, raising our power in controlling for exogenous variables.

In sum, I argue that for information goods with an interpreted component, the more two countries have imported from the same international sources, the more likely one is to begin to export to the other.  I therefore test the hypothesis that the pattern of trade in music evolves through a process of transitive closure, or more precisely,

\textsc{hypothesis} \textit{New markets for a country's music are more likely, the more transitive mediated triads the new trade relationship would complete.}

\section{Data and methods}

\subsection{Data}
The hypothesis of this paper must be tested with longitudinal data. Records of exports and imports are extracted from the United Nations Commodity Trade Database (COMTRADE 2012a). I use data classified under Standard International Trade Code Revision 2, commodity number 89832 (sound recordings), which includes pre-recorded physical sound carrying media of any form. CDs, cassettes, phonograph records, and mini-discs are all included in this category, and blank media and computer software are not. For these data, the value of an export flow is reported by each exporting country in United States dollars in ``free on board'' terms, which includes costs of the exported goods and the costs incurred in getting them to the country's border, but excludes shipping and insurance costs incurred in transporting them to the destination country.  Data availability begins in 1976, with 18 countries reporting exports to specific destinations. There is a gradually increasing trend in the number of countries reporting the destination of their exports until it reaches a maximum in 2005 and 2006 at 150 countries reporting.

Some pre-processing of data is required.  Most notably, export data from Taiwan is not available for political reasons. However, it is possible to approximate these data by examining other countries' reports of imports from ``Other Asia, nes,'' which is a term reserved in practice for Taiwan.  Some trade flows are reported with other indefinite partner designations (for example, ``World,'' or ``Areas, nes'') but unlike ``Other Asia, nes,'' these do not indicate a definite destination (COMTRADE 2012b, 2012c).  I therefore do not include these indefinite data in the present study.  

Because my hypothesis is intended to test the idea that imports of music affect the music that is created within a country, we must exclude goods that merely pass through a country on their way somewhere else.  It is therefore necessary to subtract re-exports from total exports to arrive at a figure for net exports of music recordings produced domestically (note that being produced domestically does not necessarily imply that the music was originally composed domestically).  Other authors have tended to prefer to use imports, rather than exports as the source of data on trade flows for reasons of their greater accuracy, but this approach would not have allowed the subtraction of re-exports, which is the more significant distortion of the data for my purposes. 

The hypothesis is concerned with new trade flows, but we only wish to focus on economically meaningful ones.  It is therefore necessary to dichotomize the data to represent when a trade flow crosses over some significant threshold from being 'absent' to being 'present.'  The method used to dichotomize the data could substantially affect the results of statistical models run on this data.  I therefore repeat the analyses on multiple versions of the binarized data.  The primary approach to dichotomization attempts to code a trade flow as present (a ``1''), if it is large relative to the market of the importing country.  Simply because of the difference in sizes, one million dollars' worth of exports to the United States would certainly be much less influential on the local language than the same amount of music exported to say, Belgium, or Malawi.   

To establish a suitable means of comparison, I first took available data on music market sizes (dollars spent by consumers) from the International Federation of the Phonograph Industry (IFPI 1992, 1998, 1999, 2005) and regressed them on population and GDP (with all variables log-transformed) (Table \ref{tab:SizeOfDomesticMarket}).  These variables had good predictive power ($R^{2}$ of .89) and allow for reasonable approximation of music market sizes for the many countries that do not report data through the IFPI but do report population and GDP statistics.  I then chose a threshold, $a$, such that if the value of the trade flow was greater than $a$\% of the predicted size of the importing country's music market, the flow would be coded as ``1,'' and ``0'' otherwise.  To confirm that results were not dependent on the value of $a,$ I repeated the analyses, varying $a$ from 1\% to 35\% of the estimated size of importing music market.  These results are presented in Table \ref{tab:EffectOfTieInclusionThresholdOnTransitivityParameterEstimate}, in the following section.

\begin{table}[htb]
	\centering 
	\caption{Size of domestic market}
	\begin{minipage}{\textwidth}
		\begin{tabular}{lrrrr}
		\hline\hline
	& \multicolumn{2}{c}{Full sample}&\multicolumn{2}{c}{Top 5 mkts omitted}\\
	\hline
	&Estimate	&p-value		&Estimate	&p-value\\
	\cline{2-5}
Intercept	&-15.99	&$<0.001$	&-15.464&	$<0.001$\\
ln GDP/capita	&1.23	&$<0.001$&	1.200	&$<0.001$\\
ln population	&0.988	&$<0.001$	&	0.959	&$<0.001$\\
					\\
N	&47&&			42&\\	
Adj. R sq.&	0.935&&	 	 	0.893&\\	 
\hline\hline
		\end{tabular}
	\end{minipage}
	\label{tab:SizeOfDomesticMarket}
\end{table}

In an additional series of analyses, rather than consider the relative size of trade flows, I used a dichotomization scheme based on absolute sizes.  For these runs, I used the following algorithm.  Starting with the largest reported flow by dollar value and proceeding from largest to smallest, trade flows are coded as present (as a 1), until the sum of their dollar value equals 95\% of the total global dollar value of reported trade.  The thousands of the smallest trade flows that together sum to less than five percent of total reported world trade in recorded music, are thus coded as absent (as 0).  

For each data year, I represent my trade data in a square matrix, representing all possible origin-destination pairs of countries as row-column entries. Each country is represented as a row (as the exporting country) and as a column (the importing country).  These matrices taken together can be viewed as a representation of the phenomenon to be explained: the evolving network of trade in music.

Starting with all available COMTRADE data, any country that was coded as importing or exporting in any of the data years is considered part of the study population of countries for all of years in a given set of analyses. That is, the set of countries under consideration is constant across time, and omits all countries that do not report trade in music at a high level (as defined by the inclusion algorithms described above) during the study period.  One exception to this generalization is that a number of countries were divided into several smaller countries during the study period, such as Yugoslavia.  In this case, I treat the largest (by population) ``child'' country (i.e. Serbia) as if it were the same country as the ``parent'' (Yugoslavia).  The smaller ``children'' (e.g. Croatia) are left out of statistical estimations until they become independent states by coding their entire row and column of the trade matrices with ``structural zeros''.  These encode an absent trade flow that is absent for exogenous reasons (in this case, because the country did not exist), and are not considered by the statistical model (see Ripley, Snijders, and Preciado 2011 for more detail).  The other exception is Germany, which unified the former East and West Germanies.  In this case, I treat West Germany and unified Germany as one country and East Germany as if it could no longer trade for exogenous reasons (with structural zeros).

If we wish to study when one country is likely to begin to export to the other, then this is to say that we are particularly interested in the formation of trade flows --- the change from 0s to 1s in the set of matrices representing the changing pattern of trade. To evaluate what conditions are associated with the changes from 0s to 1s, we must also take into account those 0s that remain 0s. The absence of a trade flow between a given pair of countries in the study population provides information for empirical analysis, just as the presence of a trade flow does. In this study, therefore, each entry in the matrices representing trade in music corresponds to one observation in the dependent variable.

Digital delivery of music recordings has, of course, been a major disruptive force in the music industry.  The IFPI reports that in 2006, digital music consumption amounted to only 10\% of total sales in the United States.  Although growing rapidly, it had not yet become the overwhelming influence it is today during the study period.  On the other hand, the effect of illegally copied and shared music on the volume of legitimate international trade up through 2006 is not known.  After dichotomizing the data, it does not appear to have had a dramatic effect, in that the number of trade ties continues to rise over time, even through the 2001 and 2006 data years (see Table \ref{tab:networkDescriptives}).  To confirm that there was no substantial effect on the model proposed and tested here, I run an additional model excluding those final two observation years, and compare to the full data set.  

\subsection{Non-parametric analyses}
In addition to the multivariate analyses, discussed below, I present results for non-parametric comparisons of the association between amount of common influence and changes to the trade network.  For each observation except the first (before which there is no observed history), each potential ordered pair of countries is placed into one of four categories.  One category of ordered pairs has ties that have been maintained from the previous period (``maintained ties''). The second is ties that were present in the previous period but absent in the current period (``lost ties'').  The third is ``new ties,'' and the fourth is ties that are not present in the current or previous time periods (``no ties'').  

For each matrix entry in each category, I step backwards from the observation and count the number of previous observation years in which common influences were experienced by the two countries.  In addition to descriptive statistics for each category, I employ a discrete two-sample Kolmogorov-Smirnov (KS) tests to examine whether the distribution of common influences varies by category.  The two-sample KS test is a non-parametric test for differences between the cumulative distribution functions of empirical data, especially useful when the underlying distributions are not known a priori (Massey 1951).

\subsection{Exogenous variables for multivariate models}
The empirical benchmark in describing international trade is the so-called ``gravity model'' (e.g. Anderson 1979), and I look to this tradition as a starting point for inclusion of exogenous explanatory variables. The gravity model shows that the level of trade between two countries is proportional to the product of their GDPs (analogous to the interaction of their economic 'masses') and the inverse square of the geographic distance between them. In cultural industries specifically, there are economic arguments that the size of the exporter's market is most important in determining the pattern of trade (Wildman and Siwek 1988, Dupagne and Waterman 1998, Lee and Waterman 2007).  I therefore represent the market sizes separately for the exporter (aka ``ego'') and the importer (``alter'').  In the multivariate analyses below, the effect of the covariates enters multiplicatively, so there is no substantial difference between this and the original gravity model. The gravity model literature also typically includes variables for the geographic contiguity of pairs of countries, whether they have a history of a colonial relationship together, and whether they share a spoken language.  I follow this precedent and include variables for contiguity, direct colonial tie (country A colonized country B), and indirect colonial tie (countries A and B were both colonized by country C), and whether greater than 9\% of the population of each pair of countries shares a common language.  Data for commonality of language, geodesic distance, contiguity, and colonial relationship are taken from the CEPII database of international trade statistics, available online at www.cepii.fr. Descriptive statistics for covariates and dichotomous trade networks are in Tables \ref{tab:DescriptiveStatisticsForCovariates} and \ref{tab:networkDescriptives}, respectively.

\begin{table}
	\centering
	
		\caption{Descriptive statistics for covariates}
	\begin{adjustwidth}{-1.5cm}{}	
		\begin{tabular}{lrrrrrrrr}
		\hline\hline
				&Min.	&1st Qrt.	&Median	&Mean	&3rd Qrt.	&Max.	&Variance	&unit\\
\cline{2-9}
\vspace{-5pt}\\

ln(distance)&	4.09&	8.43&	8.96&	8.77&	9.33&	9.90&	0.65&	Kilometers\\
Colony&	0&	0&	0&	0.01&	0&	1&	0.01&	indicator\\
contiguous&	0&	0&	0&	0.02&	0&	1&	0.02&	indicator\\
common language&	0&	0&	0&	0.15&	0&	1&	0.13&	indicator\\
common colonizer&	0&	0&	0&	0.11&	0&	1&	0.10&	indicator\\
Music market 1976&	1.25&	6.39&	8.06&	8.36&	10.37&	15.35&	7.33&	US\$ (2005)\\
Music market 1981&	1.52&	6.66&	8.40&	8.57&	10.63&	15.52&	7.26&	US\$ (2005)\\
Music market 1986&	1.81&	6.70&	8.49&	8.71&	10.76&	15.72&	7.18&	US\$ (2005)\\
Music market 1991&	2.21&	7.09&	8.68&	8.91&	10.82&	15.86&	6.60&	US\$ (2005)\\
Music market 1996&	2.67&	7.13&	8.83&	9.04&	11.05&	16.04&	6.71&	US\$ (2005)\\
Music market 2001&	2.97&	7.40&	9.04&	9.25&	11.18&	16.26&	6.60&	US\$ (2005)\\
Music market 2006&	3.57&	7.67&	9.43&	9.52&	11.46&	16.41&	6.58&	US\$ (2005)\\
\hline\hline
		\end{tabular}
\end{adjustwidth}
	\label{tab:DescriptiveStatisticsForCovariates}
\end{table}

\begin{table}
	\centering
		\caption{Network descriptive statistics by year}
		\begin{tabular}{llrrrrrrr}
		\hline\hline
		&&1976	&1981	&1986	&1991	&1996	&2001	&2006\\
\cline{3-9}
\multicolumn{8}{l}{at 1\% threshold}							\\
	&number of ties&	12&	259&	335&	631&	988&	1125&	1351\\
	&max outdegree&	3&	63&	71&	105&	128&	142&	136\\
	&\# with OD=1&	5&	11&	11&	8&	18&	24&	12\\
	&\# with OD=0&	182&	158	&150&	136&	111&	97&	100\\
								
\multicolumn{8}{l}{at 5\% threshold}\\
	&number of ties	&2	&93&	115&	261&	413&	472&	611\\
	&max outdegree&	1&	25&	32&	60&	80&	79&	81\\
	&\# with OD=1&	2&	10&	13&	16&	15&	22&	20\\
 	&\# with OD=0&	188&	170&	165&	151&	137&	123&	119\\
	\hline\hline
\multicolumn{8}{l}{\small \emph{Note: ``OD'' is an abbreviation for ``outdegree''}}\\
\end{tabular}
	\label{tab:networkDescriptives}
\end{table}

\subsection{Hypothesized interdependence of observations}
I am hypothesizing an effect of the existence of certain trade flows on other trade flows. Of course, we must also explain the presence or absence of the first trade flows as well, which will be explained (in part) by the presence or absence of still other trade flows, and so on. This sort of endogenous causation (trade flows depend on trade flows), implies that each observed trade flow should play dual roles in the analysis. As an import, it is a causal factor relevant to the importing country's future exports. As an export from the originating country, it is itself an outcome to be explained. The observations in the dependent variable are, by hypothesis, interdependent.  Interdependence of this sort is a clear violation of the independence assumptions of standard multivariate statistical models, but is the bread and butter of network models, in which it is known as a ``structural effect,'' in that the structure of the network endogenously influences its own subsequent structure.

\subsubsection{Competing hypotheses}
Conceptually intertwined with the main hypothesis are at least three other phenomena of interdependency. Each of these is predicted to be correlated with the transitive mediated triads of the main hypothesis, above.  These therefore act as competing hypotheses, in that they could theoretically explain observed transitive closure.

The first is the tendency toward reciprocity of already existing trade ties, as discussed above. A correlation between reciprocity and transitivity could arise in a dense area of the network.  Thus reciprocity must be controlled for.  

The second phenomenon is having a common colonizer (mentioned in ``exogenous variables,'' above).  If two countries were colonized by the same third country, it would not be surprising if they traded at an above average level with their colonizer and each other, thus creating triads.  These triads might be due to a similar process of influence to that in the main hypothesis (if the colonizer's music influenced both importing countries), but it could also be due to completely unrelated factors.  For example, such countries could have shared legal systems that support trade, a history of bilateral shipping instituted by the colonizer, or other factors in common.  

The third phenomenon is one that has been observed in international trade in differentiated industries in general: the tendency for major exporters to trade disproportionately with each other.  In the ``new'' trade theory (Krugman, 1980), countries that have successfully specialized in a variety or varieties of goods within an industry are likely export to many partner countries, selling their domestic products, as well as import from many partner countries, who make complementary products. The result is the tendency for countries with many export markets to be each other's trading partners to a greater degree than would be predicted by the numbers of their trading partners alone. In the language of network analysis, the number of outgoing links is known as ``out-degree'' and this sorting of countries within a network by the number of outgoing links is known as out-degree assortativity (hereafter ``assortativity'' for brevity).  Where a few countries have disproportionately many outgoing trade ties, and they disproportionately trade with each other, this could appear statistically like the effect of a transitive process, even if transitivity played no role in the actual formation or dissolution of ties. Literature on network dynamics suggests that assortativity and transitivity are not just difficult to distinguish, but may in fact limit the values the other can take (Serrano and Bogu\~{n}\'{a}, 2005; Holme and Zhao, 2007).  The presence of a bounded range of possible transitivity confirms that even without a true tendency to transitivity, an assortativity effect can itself produce a considerable amount of incidental transitivity.  Over time, both degree assortativity and transitivity can lead to a similar clustered pattern of ties in the network.  However, the two effects are constituted by different micro-processes.  To rule out this alternative explanation of observed transitive closure, rather than include both transitivity and assortativity effects (in which case, analogous to other multicollinearity problems, it would be difficult to determine which structural effect was responsible for the observed pattern of ties), I run analyses with finer-grained (yearly) data that include an assortativity parameter but no transitivity parameter, testing if such a specification could account for the data.

\subsection{The Siena framework}
The presence of endogenous structural effects in a model of network evolution mean that the status of each trade tie depends (in multiple ways) on the status of each other trade tie. Finding parameter estimates and evaluating their statistical significance analytically (analogous to fitting a least-squares line algebraically, for example) is simply not feasible for this type of data. Snijders and colleagues (2001, 2005, Snijders et al. 2007, Ripley, Snijders, and Preciado 2011) therefore elaborated a Markov Chain Monte Carlo type computer simulation method of arriving at maximum likelihood parameter estimates and standard errors. Snijders' longitudinal model of network evolution is known as the ``Siena'' model, which stands for ``Simulation Investigation for Empirical Network Analysis,'' now available as the package ``RSiena'' for the R statistical computing environment.
\subsubsection{Model Basics}
To convey the nature of the analytical method used in this research, I present a simplified description of the Siena model here; for a more complete and detailed description, the reader is referred to the original sources (Snijders, 2001, 2005; Snijders et al. 2007; Ripley, Snijders, and Preciado 2011). In Siena, the actual set of changes to the structure of the observed network -- in this case, changes in who exports to whom -- is considered the outcome of stochastic process, which is to be modeled. Other changes to the network could have resulted from the same underlying forces making up the stochastic process, but the modeling effort seeks parameter values that make the observed data more likely than any other combination of parameter values (thus ``maximum likelihood''). Let us denote the current state of the network as $x$, drawn from a distribution of such networks, $X$, where $x_{ijt} = 1$ when country $i$ exports to country $j$ at time $t$ and $x_{ijt} = 0$ otherwise. The probability that the network changes from one state to another, $p(X_t=x_t|X_{t-1}=x_{t-1})$ is a function of the sum of the influences from the effects in the model. The observed data are characterized in terms of ``change statistics,'' $s_{kt}$, where $t$ indicates the period and $k$ denotes which statistic is being counted, for example the number of newly reciprocated trade ties or the number of newly completed transitive mediated triads. 

The formation or dissolution of each trade tie can influence multiple change statistics simultaneously -- for example a new tie could be both a newly reciprocal tie and complete a transitive mediated triad.  In the model, these statistics are multiplied by parameters $\beta_k$ that represent the contribution of the structure in question to the likelihood of forming a tie. At a high level, the model states that the tendency for the network $X$ to take on a certain form, $x$ can be described in terms of an objective function evaluated from the point of view of nodes, $i$, and depends to various extents, $\beta_k$, on the change in the number of certain structures, $s_{ki}$ in $x$:

\[f_i(\beta,x) = \sum_{k}\beta_{k}s_{ki}(x).\]

Note that time is modeled If there is a ``true'' tendency toward the completion of transitive mediated triads (TMTs), above and beyond the amount that they appear as a by-product of other effects in the model, then the parameter for this effect, $\beta_{TMT}$ should be significantly greater than zero, and the network is more likely to evolve by adding these triads than otherwise.  

\subsubsection{Estimation algorithm}
To estimate model parameters Siena uses an iterative process of simulating networks based on the current parameter values, comparing the simulated networks to the observed data, and then updating the parameter values until model convergence is achieved.

Starting from initial parameter values, $\hat{\beta_k}$, the algorithm simulates changes to the network over time and calculates the change statistics in the simulated networks (i.e. how many reciprocated ties or transitive mediated triads were formed in the simulations over the $t$ periods) -- call these $s_{kt}^{sim}$.  These are then compared with the actual change statistics observed from the empirical data, $s_{kt}^{obs}$.  The values of $\hat{\beta_k}$ are then incremented or decremented and more simulations are conducted.  This process of simulation, comparison, and parameter update is repeated until the simulated evolution of the network looks like the observed evolution of the network, or

\[E[s_{kt}^{sim}|\hat{\beta_k}] =s_{kt}^{obs}. \]
\subsubsection{Definitions of change statistics}
The basic definitions for the change statistics for the structural variables are as follows.  The transitive mediated triads change statistic is equal to \[\sum_{h=1}^n (x_{t,h,i})(x_{t,h,j}) - \sum_{h=1}^n (x_{t-1,h,i})(x_{t-1,h,j}) \] for ${h\neq i\neq j}.$  For reciprocity, the change statistic is 
\[x_{t,j,i} - x_{t-1,j,i}\] and for out-degree assortativity, it is 
\[(\sum_{h=1}^n (x_{t,i,k}))(\sum_{h=1}^n (x_{t,j,k})) - (\sum_{h=1}^n (x_{t-1,i,k}))(\sum_{h=1}^n (x_{t-1,j,k}))\] for ${h\neq i\neq j}$. In addition to these basic definitions, there are also versions that separately measure the formation of ties and the maintenance of existing ties (see  Ripley, Snijders, and Preciado 2011 for definitions). 

\section{Results}
Non-parametric tests for a tendency toward triadic closure are positive, and in multivariate analyses the parameter for transitive mediated triads is positive and significant in all of its forms.  The main hypothesis of this paper is therefore supported.  The pattern of trade evolves through a process of transitive closure: new markets for a country's music are more likely, the more transitive mediated triads the new trade relationship would complete.  More detailed results are presented, below.

\subsection{Non-parametric comparisons}
Two sets of unconditional comparisons were performed (Table \ref{tab:UnconditionalComparisonsOfCommonHistoryByTieCategory}).  The first compares all ordered pairs of countries at all time points that are not connected by a trade tie.  Within these, I compare the set that go on to form a trade tie to the set that remains without a tie in the next observation.  The other set considers ordered country pairs that are tied and compares those that lose that tie to those that maintain it in the next observation. 

Kolmogorov-Smirnov tests show that the sets that had ties at the second observation (those that created a new tie or maintained a tie) had experienced significantly more social influence in common than those that did not have ties at the second observation (those that lost ties or which maintained a lack of tie). The only comparison that was not statistically significant was the length of individual influences held in common (as opposed to the total of all influences or the number of influences) at the 5\% threshold for tie inclusion.  All other comparisons at both the 1\% and 5\% thresholds were significant at the $p<0.001$ level.  Pairs of countries that formed ties or maintained ties had more social influence in common than pairs of countries that lost or did not form ties.

\begin{table}
	\caption{Non-parametric comparisons of common history by tie category }
	\centering
	\begin{adjustwidth}{-1cm}{}	
	\begin{minipage}{\textwidth}
		\begin{tabular}{llrrrr}
		\hline\hline
			 \multicolumn{6}{c}{1\% threshold}\\
			\hline
 	
		&&maintained ties&	lost ties&	 	new ties&	no ties\\
		\cline{3-4}
		\cline{5-6}
\multicolumn{6}{l}{Descriptive statistics}					\\
	&mean total common influence (years)	&20	&15.9		&10.6	&4.1\\
	&mean duration of a common influence 	&8.8	&8.8		&8.2	&7.8\\
	&mean number of common influences	&2.3	&1.8	&1.3	&0.5\\
	&n observed	&4484	&2216		&4884	&365756\\
\multicolumn{6}{l}{K.S. test statistics and p-values}					\\
	&total influence	&\multicolumn{2}{r}{$0.06, p < 0.0001$}&\multicolumn{2}{r}{$0.14, p < 0.0001$}\\
	&length of single influences in common&\multicolumn{2}{r}{$	0.03, p < 0.0001$}&\multicolumn{2}{r}{$		0.06, p < 0.0001$}\\
 	&number of influences in common&\multicolumn{2}{r}{$	0.07, p < 0.0001	$}&\multicolumn{2}{r}{$ 	0.14, p < 0.0001$}\\
\\
\hline
			 \multicolumn{6}{c}{5\% threshold}\\
\hline
		&&maintained ties&	lost ties&	 	new ties&	no ties\\
			\cline{3-4}
		\cline{5-6}
\multicolumn{6}{l}{Descriptive statistics}					\\
	&mean total common influence (years)&	6.3&	4.6	&	3.4&	1\\
	&mean duration of a common influence	&8.1	&8		&7.6	&7.2\\
	&mean number of common influences	&0.8	&0.6&		0.5	&0.1\\
	&n observed	&1742	&970		&2184	&372444\\
\multicolumn{6}{l}{K.S. test statistics and p-values}					\\
	&total influence	&\multicolumn{2}{r}{$0.05, p < 0.001	$}&\multicolumn{2}{r}{$ 	0.12, p < 0.001$}\\
	&length of single influences in common&	\multicolumn{2}{r}{$0.03, p = 0.200	$}&\multicolumn{2}{r}{$ 	0.08, p < 0.001$}\\
 	&number of influences in common&\multicolumn{2}{r}{$	0.06, p < 0.001$}&\multicolumn{2}{r}{$ 		0.12, p < 0.001$}\\
\hline\hline
\multicolumn{6}{p{5.5in}}{\textit{Note: The Kolmogorov-Smirnov test statistic is the maximum vertical difference between the empirical cumulative distribution functions. The p-value is the probability that the two samples are drawn from the same population.  Comparisons are made between maintained and lost ties and between new ties and no ties.}}
\vspace{-5pt}\\
		\end{tabular}
		\vspace{-30pt}\\
\end{minipage}
\end{adjustwidth}
	\label{tab:UnconditionalComparisonsOfCommonHistoryByTieCategory}
\end{table}

The descriptive statistics in Table \ref{tab:UnconditionalComparisonsOfCommonHistoryByTieCategory} also give a sense of the time scale at work. Broadly speaking, social influence can be seen to operate after one or two observations (between 5 and 10 years) on average.  Those that formed ties had on average 10 years of common influence, which would be equivalent to two sources of common influence for one observation period, or one source for two. 

\subsection{Multivariate models}
\label{effectsizes}
Reported parameter estimates from the multivariate Siena models can be understood as log odds ratios that weight the contribution of the covariate that they modify to the probability that a trade tie exists. The magnitude of an effect on the probability of a tie attributable to a difference in covariate values is $e^{\hat{\beta}(\Delta s_k)}$, where $\Delta s_k$ denotes the difference in covariates, and $\hat{\beta}$ the estimated parameter.  For example, if we wanted to compare the difference in tie probability between contiguous and non-contiguous countries (a dummy variable), $\Delta s_k$  would be 1, and the effect magnitude would be $e^{\hat{\beta}}$. If we wanted to compare how much more likely a tie was to originate between countries one standard deviation closer than the mean distance between all countries, we would plug in -0.804 (minus one standard deviation (see Table \ref{tab:DescriptiveStatisticsForCovariates} for descriptive statistics)) for $\Delta s_k$.  As such, the size of the effect depends on the range and variability of each individual variable, as well as the estimated parameter. Parameters cannot be directly compared to those estimated for other variables without reference to covariate values.

The main results for this study are presented in Table \ref{tab:ModelsEstimatedFrom19762006DataObservationsTakenAtFiveYearIntervals}.  This series of models is estimated from observations taken at 5 year intervals from 1976 to 2006.  The first column represents the basic model, with all variables parametrized in the ``evaluation'' form.  The evaluation form means that parameters weight the importance of the covariates in contributing to the presence of a trade tie, without distinguishing between new ties and existing ties.  The second model divides each variable into ``creation'' (tendency to form new ties) and ``endowment'' (tendency to maintain existing ties) parametrization.  When the absolute value of the parameter estimate is more than twice the standard error, the parameter can be considered significant at the $p<0.05$ level.

\begin{table}
	\centering
		\caption{Models estimated from 1976-2006 data, observations taken at five year intervals}
	
		\begin{tabular}{llrrrr}
		\hline\hline
		&&est.&	s.e.	&est.	&s.e.\\
		\cline{3-6}
\multicolumn{2}{l}{outdegree (density)}&	-5.01&	0.08&	-5.50&	0.09\\
\multicolumn{2}{l}{reciprocity}&	0.18&	0.06		\\
	&endowment&&			&2.09	&0.25\\
	&creation	&&		&-0.29	&0.13\\
\multicolumn{2}{l}{transitive med triads}&	0.26&	0.01\\		
	&endowment&&			&0.11	&0.04\\
	&creation	&&		&0.33	&0.03\\
\multicolumn{2}{l}{$ln$(distance)}&	-0.47&	0.02		\\
	&endowment&&			&-0.73	&0.07\\
	&creation&&			&-0.31	&0.04\\
\multicolumn{2}{l}{colonial tie (direct)}&	0.58	&0.06	\\	
	&endowment	&&		&0.35&	0.21\\
	&creation	&&		&0.67	&0.14\\
\multicolumn{2}{l}{contiguous}&	0.28	&0.06		\\
	&endowment	&&		&0.55	&0.23\\
	&creation		&&	&0.25	&0.12\\
\multicolumn{2}{l}{common language}&	0.76	&0.04	\\	
	&endowment	&&		&2.01	&0.14\\
	&creation		&&	&0.00	&0.08\\
\multicolumn{2}{l}{common colonizer}&	0.26	&0.07\\		
	&endowment	&&		&2.57	&0.28\\
	&creation		&&	&-0.83	&0.16\\
\multicolumn{2}{l}{music market  alter}&	-0.02	&0.01\\		
	&endowment	&&		&0.05	&0.03\\
	&creation		&&	&-0.06	&0.02\\
\multicolumn{2}{l}{music market ego}&	0.62	&0.01\\		
	&endowment	&&		&-1.29	&0.05\\
	&creation		&&	&2.71	&0.07\\
					
\multicolumn{2}{l}{rate 1976-1981}&	30.92	&4.43	&50.10	&7.13\\
\multicolumn{2}{l}{rate 1981-1986}&	10.73	&1.18	&17.61	&2.02\\
\multicolumn{2}{l}{rate 1986-1991}&	43.91	&4.97	&68.54	&7.68\\
\multicolumn{2}{l}{rate 1991-1996}&	39.66	&3.15	&56.90	&4.54\\
\multicolumn{2}{l}{rate 1996-2001}&	27.10	&1.60	&37.42	&2.44\\
\multicolumn{2}{l}{rate 2001-2006}&	23.99	&1.33	&32.43	&1.98\\
 	 	 	 	 	
					\hline\hline
\multicolumn{6}{l}{\emph{Note: ties coded as present at the 1\% threshold}}
		
		\end{tabular}
	\label{tab:ModelsEstimatedFrom19762006DataObservationsTakenAtFiveYearIntervals}
\end{table}

  To give a sense of the importance of the main result, Table \ref{tab:MagnitudeOfCommonInfluenceEffect} shows calculated effect magnitudes for different combinations of geographic distance and number of common influences.  This comparison shows that a single common influence is estimated to have a greater positive correlation with the probability of a new exporting relationship than would a hypothetical change in geographic distance of one standard deviation on the log scale.  
	
	Unlike the other variables in the model, the parameter for common influences (transitive mediated triads) may enter multiplicatively if there is more than one common influence.  Each common influence multiplies the odds of tie formation by about 1.39.  So, two influences in common would mean the odds of new ties is $1.39^2$, and three influences means that new ties are $1.39^3$ times more likely than without any influences in common.  Considering the observed numbers of influences in common, this could represent a very large effect.  For example, in these data more than 800 ordered country pairs had 8 or more common influences, representing an increase in new tie probability of at least 13.9 times the set of countries with no influences in common (see Table \ref{tab:MagnitudeOfCommonInfluenceEffect}).

\begin{table}
	\centering
	
		\caption{Magnitude of common influence effect}
		\begin{minipage}{\textwidth}
		\centering
		\begin{tabular}{lrrrrr}
\hline\hline
		distance (miles)&	\multicolumn{5}{c}{Number of common influences}\\
	&0	&1	&2	&3	&4\\
	\cline{2-6}
12456 (max)	&0.70&	0.98&	1.36&	1.89&	2.63\\
8866 (+1 sd)&	0.78&	1.08&	1.51&	2.10&	2.92\\
3983 (mean)	&1	&1.39&	1.93&	2.69&	3.74\\
1790 (-1 sd)&	1.28	&1.78&	2.48&	3.45&	4.80\\
804 (-2 sd)&	1.64	&2.28&	3.18&	4.42&	6.15\\
37 (min)	&4.27&	5.93&	8.25&	11.48&	15.97\\
\hline\hline
\multicolumn{6}{p{3.5in}}{\textit{Notes: Indications such as (+1 sd) refer to an increase of one standard deviation on the log scale.  Table entries are odds of a new tie being created, relative to countries with no common influences and countries 3983 miles apart (the mean of the distance variable on the log scale).}}

		\end{tabular}
		\vspace{0pt}\\
\end{minipage}
	\label{tab:MagnitudeOfCommonInfluenceEffect}
\end{table}

\subsubsection{Additional Results} Plugging parameter estimates into $e^{\hat{\beta}(\Delta s_k)}$ to get effect magnitudes (see beginning of section \ref{effectsizes}, above), other factors in the formation of new ties are as follows. New ties that reciprocate existing ties are only 75\% as likely as non-reciprocal ties, ruling out reciprocity as a competing explanation for the finding of a tendency to transitivity. Geographic distance is associated with fewer new ties; countries that are a single standard less distant on the log scale are 1.28 times more likely to form a trade tie.  Contiguous countries are an additional 1.28 times more likely to form a tie than non-contiguous countries.  Unsurprisingly, a past direct colonial relationship makes a new tie almost twice as likely as otherwise, but interestingly, countries that were colonized by the same colonizer were less than half as likely (44\%) to form new ties than those that did not share a colonizer.  This rules out the competing hypothesis of common colonizers for explanation of new ties that complete triadic structures in the pattern of trade.  The effect with the greatest magnitude was the estimated size of the music market: doubling the size of a music market multiplies the probability of creating a new outgoing tie by more than 6.5. There was a very slight decrease in the probability of exporting to countries with larger music markets.  Doubling market size of the destination country is associated with a reduction of new tie probability by about 4\%.  

The correlates of the maintenance of existing ties are somewhat different.  Reciprocal ties are about 8 times more likely to be maintained than non-reciprocal ties.  Each common influence multiplies the odds of tie maintenance by about 1.11 (a much smaller effect than in tie formation).  For geographic distance, a single standard deviation on the log scale has a greater effect for maintaining ties --- about 1.44 times more likely for more proximate countries, with an additional 1.73 times the probability of maintaining ties for contiguous countries. A direct colonial tie makes maintaining ties about 1.42 times as likely, but countries that were colonized by the same colonizer are more than 13 times more likely to maintain their existing music trade.  Having a language in common had no significant correlation with the creation of ties, but it is associated with maintaining existing ties around 7.4 times more of the time. In contrast to tie formation, countries with larger music markets were much less likely to maintain their ties.  Doubling music market size was associated with ties being maintained 58\% less of the time.
	
\subsection{Robustness checks}
\subsubsection{Threshold for tie inclusion}
In addition to the results already presented, I conducted many robustness checks. 

I varied the threshold used to code a trade flow as present or absent from 1\% to 33\% of the importing country's estimated music market size (Table \ref{tab:EffectOfTieInclusionThresholdOnTransitivityParameterEstimate}).  The estimated parameter for transitive mediated triads increases steadily from .26 to 1.66 as the threshold increases. This is not just reassuring but indeed strengthens the argument that social influence is occurring, because larger trade flows could be interpreted to exert more influence than small trade flows.

\begin{table}
	\centering
	\caption{Effect of tie-inclusion threshold on transitivity parameter estimate}
\begin{minipage}{\textwidth}
\centering
		\begin{tabular}{lrrrrrrrrr}
\hline\hline
			&\multicolumn{9}{c}{Threshold}\\	
			&1\%&	5\%&	9\%&	13\%&	17\%&	21\%&	35\%&	29\%&	33\%\\
\cline{2-10}
Estimate&0.26	&0.50&	0.68&	0.73&	0.92&	1.08&	1.43&	1.43&	1.66\\
Standard error 	&0.01&	0.04&	0.07&	0.13&	0.16&	0.19&	0.25&	0.29&	0.39\\
\hline\hline
\multicolumn{10}{p{4.6in}}{\textit{Note: the tie inclusion threshold is the percentage of the estimated size of a music market above which an incoming tie is coded as 1 and below which, 0.}}

		\end{tabular}
		\vspace{0pt}\\
		\end{minipage}
	\label{tab:EffectOfTieInclusionThresholdOnTransitivityParameterEstimate}
\end{table}
\subsubsection{Tests of other models specification choices}
Both economic theory and network science provide reasons to suspect that assortativity may be responsible for observed transitive closure.  On these data at least, that suspicion is not justified.  I re-estimated models on 15 one year periods and specified with an out-degree assortativity parameter, but with the transitivity parameter fixed at zero.  Both linear and square root assortativity specifications were considered.  Rao score tests of the null hypothesis that the parameter for transitive mediated triads is equal to zero (because transitivity is accounted for by assortative processes) were conducted, and the null hypothesis was rejected (see Table \ref{tab:ScoreTestOfModelRestrictedSTTransitiveMediatedTriadParameterIsFixedAtZero}).  That is, assortativity cannot explain the level of transitivity that we observe in the data.

\begin{table}
	\centering
		\caption{Score test of model restricted s.t. transitive mediated triad parameter is fixed at zero}
		\begin{adjustwidth}{-.2cm}{}	
	\begin{minipage}{\textwidth}
	\centering
		\begin{tabular}{lrrr}
		\hline\hline
		&no assortativity& assortativity&assortativity$^{1/2}$\\
		\cline{2-4}
		Estimated assortativity $\beta$&-&0.0005&0.0068\\
		Estimated assortativity s.e.&-&31.607&0.0024\\
		Test statistic, $c$&1291.4&985.9&1372.5\\
		$p-$value&$<0.0001$&$<0.0001$&$<0.0001$\\
				\hline\hline
	\end{tabular}
		\vspace{0pt}\\
		\end{minipage}
		\end{adjustwidth}
	\label{tab:ScoreTestOfModelRestrictedSTTransitiveMediatedTriadParameterIsFixedAtZero}
\end{table}

The following paragraphs report a series of additional robustness checks for which I report the estimate for the transitive mediated triads parameter. All of these parameter estimates should be compared to 0.26, which was the main effect reported in Table 5.

To confirm that the choice of 5-year intervals did not unduly influence the results, I repeated the analysis with lags of three, four, six and seven years.  The parameters from these analyses were all at least as high as the one estimated on 5-year intervals., with parameter estimates for lags of 3, 4, 6, and 7 years equal to 0.30, 0.28, 0.27, and 0.30, respectively.  

Additionally, as mentioned above, it is likely that some countries in the sample simply do not have the industrial infrastructure to export.  So although social influence through importing may be occurring, we would not see evidence of it in their exports. To test the robustness of this specification, I also ran a series of models including only the subset of countries which export to at least one other country during the study period. I also varied the time lag between observation years from one to seven years.  These models show substantially the same results as the models with the full sample: including only those with at least one export market produced a parameter estimate of 0.36.

The set of models run on ties coded according to their absolute, rather than relative, magnitude also agreed with the main results. After discarding the smallest trade flows in absolute terms such that their combined total is only 5\% of aggregate music trade, the parameter estimate for transitive mediated triad creation was 0.28.  

The model which omitted the years which may have been subject to the distorting influence of digital distribution of music recordings produced similar findings to the full model, with the effect of common influences, distance, and colonial relationship estimated to have slightly stronger effects than the full model (transitive mediated triad parameter estimated at 0.35).  Digital distribution may indeed have begun to have a real effect on the estimated models by moving some of the trade outside of the data set, but this distortion does not substantially affect the conclusions of the analysis.  Indeed, the relative size of the effect is stronger when we omit the years when the data may be distorted in this way.

\subsubsection{Other industries}
Chu-Shore (2010) presents a cross-sectional analysis of the network structure of multiple industries.  Perhaps due to the limits of cross-sectional data, Chu-Shore (2010) drew different conclusions from those in this paper, but they can be understood complement the present study.  As a sort of falsification exercise, we could imagine what a similar analysis would find if it were conducted on commodities or purely utilitarian industries in which social influence were implausible, such as exports of coal, or gear cutters.  In those industries, Chu-Shore found little to no evidence of a transitivity effect.  In other goods in which subjective valuation plays a strong role, such as books and furniture, there was a strong transitivity effect, lending further support to the argument here.

As one example of such a falsification exercise conducted on longitudinal data, I repeated the analysis of this paper on trade in coal (specifically Anthracite, commodity number 3221 in the SITC revision 2 system).  As expected, there was no tendency toward transitive mediated triads: the estimated parameter was -0.50 and not significant. Rather than triadic closure, trade in coal is characterized by hubs.  To converge successfully, this model required the addition of the ``out-degree activity'' parameter (tendency toward degree dispersion, which we would expect in an industry driven by comparative advantage due to natural endowments). The out-degree activity parameter for coal was positive and significant at 0.51.  In contrast, for music data, an added out-degree activity parameter was estimated at only 0.01 and the estimate for transitive mediated triads was unchanged at 0.26.  

\subsubsection{Other network models of international trade}
Many other notable network-based studies of international trade have been carried out in recent years (McNerney, Fath, \& Silverberg, 2013; Riccaboni \& Schiavo, 2014; Ward, Ahlquist \& Rozenas, 2013), and while most are not directly relevant to the present results, Bosker and Westbrock (2014) and Chaney (2014) both provide supply-side reasons to predict triadic closure in international trade flows. Both papers study aggregate trade (Bosker and Westbrock (2014) does distinguish between final and intermediate goods), rather than trade in a single industry. Setting aside the above evidence (Chu-Shore, 2010) that network structure depends on the information-intensiveness of the industry, it would be reassuring to rule out these explanations for transitivity.  

Conveniently, the two papers both predict different forms of triadic closure than the transitive mediated triad I 
observe in music trade.  Bosker and Westbrock's (2014) model implies closure as follows.  Consider country $j$, which 
imports from country $i$, and assume the existence of country $k$. Further assume that country $k$ is geographically 
close to $i$ and $j$ such that some level of trade is likely.  In order to produce goods for export to $j$, country $i
$'s demand for intermediate goods from country $k$ is expected to increase.  In other words, given tie $x_{ij}$, they 
predict the formation or strengthening of tie $x_{ki}$. This would amount to tendencies toward transitive closure, of 
the non-''mediated'' type (given the existence $x_{kj}$) and/or three-cycles (given $x_{jk}$).  If we further assume 
the existence of $x_{ji}$ (the reciprocal tie to $x_{ij}$), they would also predict transitive mediated triads.  In 
sum Bosker and Westbrock (2014) predict transitive (non-mediated) closure, three cycles and possible a combination of 
reciprocity and transitive mediated triads.  

Chaney (2014) posits a process of trade partner discovery very similar to the models of Vazquez (2003) and Saram{\"a}ki \& Kaski (2004), in which trading partners can be found either at random, or by transitive closure.  The actual direction of the trade flow is not determined by the transitive closure process, and multiple types of closed triads are consistent with the model.  Chaney's model would predict the existence of transitive mediated triads, transitive (non-mediated) triads, and three-cycles.

The model I present here only predicts a significant tendency to transitive mediated triads, and not the other results.  Additional robustness checks indicate no significant tendency to either three-cycles (parameter estimate $= -1.40, s.e.= 0.75$) or transitive non-mediated triads (parameter estimate $= 0.04, s.e. = 0.12$).   The results above already indicate a tendency against reciprocity, providing further evidence against Bosker and Westbrock (2014).  These negative results argue strongly against the supply-side models of Bosker and Westbrock (2014) and Chaney (2014) in explaining the current results.  This is not surprising, given that those papers study aggregate trade, and these results in no way call into question the validity of their findings.

\section{Discussion}
\subsection{Summary and interpretation}
Consuming foreign products is correlated with the future ability to sell in new markets --- but only in those markets that were also influenced by those same foreign products.   Markets come to have compatible demand for information goods through exposure to the same social influences: for interpreted information goods, between-market trade evolves through a process of transitive closure.

By focusing on triadic structures in the pattern of trade, we can see how social influence complements homophily in the formation of new markets for a country's (or other ``home industry's'') products.  Given the highly subjective and varied nature of music recordings, we can safely assume that all ties in the network are homophilous.  Taste in the importing country has to be compatible with the music being offered by exporters, or there would be no trade.  However, what homophily cannot explain is how the pattern of trade will change and which new ties will come into being.  A process of social influence predicts triadic closure by providing common reference points for markets that previously lacked them.  Quite analogously to the everyday experience of individual people, for markets, common experience increases the probability of new relationships.

\subsection{Conclusion}The present study has considered the question of how a local industry for interpreted information goods begins to sell their products in a new market, focusing on whether social influence from outside markets would help or hurt this effort.  The results show strong positive effects of social influence from imports in developing new export markets, but only in those countries that have experienced the same social influences: the pattern of trade in music evolves through a process of transitive closure.  More fundamentally, the results support the conclusion that the more social influence two markets have in common, the more mutual demand those markets will have for each other's interpreted goods.  Adam Smith's (1776) oft-quoted maxim is ``the division of labor is limited by the extent of the market.'' In a sense, this research attempted to consider an even more fundamental question: what is the extent of the market limited by? The answer from these data is that for interpreted information goods, the extent of the market is limited by the amount of social influence in common. 
	
\section*{References}
\begin{list}{}{%
\setlength{\topsep}{0pt}%
\setlength{\leftmargin}{0.1in}%
\setlength{\listparindent}{-0.1in}%
\setlength{\itemindent}{-0.1in}%
\setlength{\parsep}{\parskip}%
}%
\item[]
African Union. 2005. ``Nairobi Plan Of Action For Cultural Industries In Africa'' First Ordinary Session Of AU Conference Of Ministers Of Culture Nairobi, Kenya

Amsden, A. H. (2001). \emph{The Rise of ``the Rest:'' Challenges to the West from Late-Industrializing Economies}. Oxford University Press.

Anderson, J.E. 1979. ``A Theoretical Foundation For The Gravity Equation'' \emph{American Economic  Review} 69:1 pp 106-116

Aral, S., Muchnik, L., and Sundararajan, A. 2009. ``Distinguishing influence-based contagion from homophily-driven diffusion in dynamic networks'' \emph{PNAS }2009 106 (51) 21544-21549

Burt, R. S. 2004. ``Structural Holes and Good Ideas'' \emph{American Journal of Sociology} 110:2, pp. 349-399

Carlile, P. 2002. ``A Pragmatic View Of Knowledge And Boundaries: Boundary Objects In New Product Development'', \emph{Organization Science} 13:4, pp 442-455

Chu-Shore, J, 2010. "Homogenization and Specialization Effects of International Trade" \emph{World Development} 38:1, pp. 37-47

Cohen, W. M. and Levinthal, D.A. 1990. ``Absorptive Capacity: A New Perspective on Learning and Innovation.'' \emph{Administrative Science Quarterly} 35(1, Special Issue: Technology, Organizations and Innovation): 128-152.

COMTRADE. 2012a. http://comtrade.un.org/  Accessed 3/12

COMTRADE. 2012b. http://comtrade.un.org/  ``Areas not elsewhere specified'' Knowledgebase article, Accessed 3/12

COMTRADE. 2012c. http://comtrade.un.org/  ``Taiwan, Province of China Trade data'' Knowledgebase article Accessed 3/12

Dupagne, M., and Waterman, D. 1998.  Determinants of U.S. Television Fiction Imports in Western Europe. J\emph{ournal of Broadcasting and Electronic Media}, 42(2), 208-220

Hausmann, R., Hwang, J., \& Rodrik, D. (2007). What you export matters. \emph{Journal of Economic Growth}, 12(1), 1-25.

Henry, R. 2001. ``Caribbean Music In The International Context: Dancing To The Beat Of A Small Place.'' Music Industry Workshop, Third United Nations Conference On The Least Developed 
Countries, Proceedings Of The Youth Forum (pp. 59-64). Brussels: United Nations.

Hidalgo, C. A., Klinger, B., Barabasi, A. L., \& Hausmann, R. (2007). The product space conditions the development of nations. \emph{Science}, 317(5837), 482-487.

Holme, P. and Zhao, J. 2007. ``Exploring the Assortativity-Clustering Space of a Network's Degree Sequence'' \emph{Physical Review E} 75:4

Hoskins, C., Mirus, R. and Rozeboom, W. 1989. ``U.S. Programs In The International Market: Unfair Pricing?'' \emph{Journal Of Communication}, 39(2), 55-75.

Huang, R.R. 2007. ``Distance and Trade: Disentangling Unfamiliarity Effects And Transport Cost Effects'' \emph{European Economic Review} 51:1 pp 161-181

IFPI. 1992, 1998, 1999, 2005. \emph{The Record Industry in Numbers}, London: International Federation of the Phonograph Industry (IFPI)

Krugman, P. 1980. Scale Economies, Product Differentiation, And The Pattern Of Trade. \emph{American Economic Review}, 70(5), 950-959.

Krugman, P. 1991. ``Increasing Returns and Economic Geography'' \emph{Journal of Political Economy} 99:3 pp 483-499

Lakoff, G. 1987.  \emph{Women, Fire and Dangerous Things: What Categories Reveal about the Mind} Chicago: University of Chicago Press

Lee, F. L. F. 2006. ``Cultural Discount and Cross-Culture Predictability: Examining the Box Office Performance of American Movies in Hong Kong'' \emph{Journal of Media Economics} 19.4 

Lee, S-W., and Waterman, D. 2007. ``Theatrical Feature Film Trade In The United States, Europe, And Japan Since The 1950s: An Empirical Study Of The Home Market Effect.'' \emph{Journal of Media Economics}, 20(3),167-188.

Linder, S.B. 1961. \emph{An Essay on Trade and Transformation}. New York: John Wiley \& Sons.

Maines, D.R. 1977. ``Social Organization and Social Structure in Symbolic Interactionist Thought'' \emph{Annual Review of Sociology} 3: 235-259

Malmberg, A., and Maskell, P. 1997. ``Towards an Explanation Of Regional Specialization And Industry Agglomeration'' \emph{European Planning Studies} 5.1 (1997).

Massey, F. J. 1951. ``The Kolmogorov-Smirnov Test for Goodness of Fit'' \emph{Journal of the American Statistical Association} Vol. 46, No. 253 pp. 68-78

McNerney, J., Fath, B. D., \& Silverberg, G. (2013). Network structure of inter-industry flows. \emph{Physica A: Statistical Mechanics and its Applications}, 392(24), 6427-6441.

Park, S. 2006. ``China's Consumption of Korean Television Dramas: An Empirical Test of the 'Cultural Discount' Concept.''\emph{ Korea Journal }44.4 (winter): 265-290.

Parker, G. G., \& Van Alstyne, M. W. (2005). Two-sided network effects: A theory of information product design. \emph{Management Science}, 51(10), 1494-1504.

Polanyi, M. 1966. \emph{The Tacit Dimension}. New York: Doubleday

Polanyi, M. and Prosch, H. 1977.\emph{ Meaning.}  Chicago:  University of Chicago Press

Rauch, J.E. and Trindade, V. 2009. ``Neckties in the Tropics: A Model of International Trade and Cultural Diversity''. \emph{Canadian Journal of Economics/Revue Canadienne d'Economique}, 42:3, pp. 809-843,

Ricardo, D. 1891. \emph{Principles of Political Economy and Taxation} London: George Bell and Sons

Riccaboni, M., \& Schiavo, S. (2014). Stochastic trade networks. \emph{Journal of Complex Networks}, 2(4), 537-556.

Ripley, R, Snijders, T.A.B., and Preciado, P. 2011.\emph{ Manual for RSiena} University of Oxford: Department of Statistics; Nuffeld College

Salganik, M.  J., Dodds, P.S., and Watts, D.J.  2006. ``Experimental Study of Inequality And Unpredictability In An Artificial Cultural Market.'' \emph{Science}, 311, 854-856.

Salganik, M.J. and Watts, D.J. 2009. ``Web-Based Experiments for the Study of Collective Social Dynamics in Cultural Markets'' \emph{Topics in Cognitive Science} 1 439-468

Saram{\"a}ki, J., \& Kaski, K. (2004). Scale-free networks generated by random walkers. Physica A: Statistical Mechanics and its Applications, 341, 80-86.

Schultz, G.G. 1999. ``International Trade in Art'' \emph{Journal of Cultural Economics} 23:109-136.

Serrano, M.A. and Bogu\~{n}\'{a}, M. 2005. ``Tuning Clustering in Random Networks with Arbitrary Degree Distributions'' \emph{Physical Review E} 72:3

Smith, A. 1776. \emph{An Inquiry Into The Nature And Causes Of The Wealth Of Nations} Oxford: Clarendon Press

Snijders, T.A.B. (2001). ``The statistical evaluation of social network dynamics.'' \emph{Sociological Methodology}, 31(1), 361-395.

Snijders, T.A.B. 2005. ``Models for Longitudinal Network Data'' Chapter 11 in P. Carrington, J. Scott, \& S. Wasserman (Eds.), \emph{Models and Methods in Social Network Analysis}. New York: 
Cambridge University Press

Snijders, T.A.B., Steglich, C.E.G., Schweinberger, M., and Huisman, M. 2007. \emph{Manual for Siena Version 3}. University of Groningen, ICS, Groningen. University of Oxford, Department of 
Statistics, Oxford.

Storper, M. 1992.  ``The Limits to Globalization: Technology Districts and International Trade'' \emph{Economic Geography} 68:1, pp 60-93

Storper, M., and Christopherson, S. 1987. ``Flexible Specialization and Regional Industrial Agglomerations: The Case of the U.S. Motion Picture Industry.'' \emph{Annals of the Association of 
American Geographers}, 77(1), 104-117.

UNESCO. 2005. Convention on the Protection and Promotion of the Diversity of Cultural Expressions. Paris: UNESCO.

Vazquez, A. (2003). Growing network with local rules: Preferential attachment, clustering hierarchy, and degree correlations. \emph{Physical Review E}, 67(5), 056104.

Ward, M. D., Ahlquist, J. S., \& Rozenas, A. (2013). Gravity's rainbow: A dynamic latent space model for the World Trade Network. \emph{Network Science}, 1(01), 95-118.

Wildman, S. S., \&  Siwek, S. E. 1988.  \emph{International Trade in Films and Television Programs}. Cambridge, MA: Ballinger.

Witt, Ulrich. 2001. ``A theory of wants and the growth of demand'' \emph{Journal of Evolutionary Economics}, vol. 11(1), pages 23-36.

World Trade Organization Council for Trade in Services. 1998. Audiovisual Services; Background Note by the Secretariat. WTO document 98-2437, available at www.wto.org/english/tratop e/serv e/w40.doc

\end{list}

\end{document}